\documentclass[aps,prl,reprint,floatfix,superscriptaddress]{revtex4-2}
\usepackage{blindtext}
\usepackage{graphicx}
\usepackage{braket}
\usepackage{color}
\usepackage{siunitx}
\usepackage{amsmath}
\usepackage{tabularx}
\usepackage{xspace}
\usepackage{amssymb}
\usepackage{statmath}
\usepackage{bm}
\usepackage{color}

\def\mem#1#2#3{  \left\langle #1 \left\vert  #2 \right\vert #3 \right\rangle   }
\newcommand{\fm}[1]{\ifmmode#1\else$#1$\fi}

\newcommand{\HCI}[2]{\fm{\text{#1}^{#2+}}\xspace}
\newcommand{\Ca}[0]{\HCI{Ca}{14}}
\newcommand{\Be}[0]{\HCI{Be}{}}

\newcommand{\threePone}[0]{\fm{{}^3\text{P}_1}\xspace}
\newcommand{\threePzero}[0]{\fm{{}^3\text{P}_0}\xspace}

\begin{document}
%TC:ignore    

	\title{Excited-state magnetic properties of carbon-like \HCI{Ca}{14}}

 \author{Lukas J. Spieß}
	\author{Shuying Chen}
    \email{shuying.chen@quantummetrology.de}
	\author{Alexander Wilzewski}
	\author{Malte Wehrheim}
	\affiliation{Physikalisch-Technische Bundesanstalt, Bundesallee 100, 38116 Braunschweig, Germany}

    \author{Jan Gilles}
	\author{Andrey Surzhykov}
	\affiliation{Physikalisch-Technische Bundesanstalt, Bundesallee 100, 38116 Braunschweig, Germany}
	\affiliation{Institut für Mathematische Physik, Technische Universität Braunschweig, Mendelssohnstraße 3, 38106 Braunschweig, Germany}
	
	\author{Erik Benkler}
	\author{Melina Filzinger}
  	\author{Martin Steinel}
	\author{Nils Huntemann}
	\affiliation{Physikalisch-Technische Bundesanstalt, Bundesallee 100, 38116 Braunschweig, Germany}
 
    \author{Charles Cheung}
    \affiliation{Department of Physics and Astronomy, University of Delaware, Newark, Delaware 19716, USA}
    
    \author{Sergey G. Porsev}
    \affiliation{Department of Physics and Astronomy, University of Delaware, Newark, Delaware 19716, USA}
    
    \author{Andrey I. Bondarev}
    \affiliation{Helmholtz-Institut Jena, 07743 Jena, Germany}
    \affiliation{GSI Helmholtzzentrum f\"ur Schwerionenforschung GmbH, 64291 Darmstadt, Germany}
    
    \author{Marianna S. Safronova}
    \affiliation{Department of Physics and Astronomy, University of Delaware, Newark, Delaware 19716, USA}

    \author{José R. {Crespo López-Urrutia}}
	\affiliation{Max-Planck-Institut für Kernphysik, Saupfercheckweg 1, 69117 Heidelberg, Germany}
	
	\author{Piet O. Schmidt}
	\email{piet.schmidt@quantummetrology.de}
	\affiliation{Physikalisch-Technische Bundesanstalt, Bundesallee 100, 38116 Braunschweig, Germany}
	\affiliation{Institut für Quantenoptik, Leibniz Universität Hannover, Welfengarten 1, 30167 Hannover, Germany}
 
	\begin{abstract}
	
        We measured the $g$-factor of the excited state \threePone in \HCI{Ca}{14} ion to be $g = 1.499032(6)$ with a relative uncertainty of $4\times10^{-6}$. The magnetic field magnitude is derived from the Zeeman splitting of a \Be ion, co-trapped in the same linear Paul trap as the highly charged \HCI{Ca}{14} ion. 
        Furthermore, we experimentally determined the second-order Zeeman coefficient $C_2$ of the \threePzero\ - \threePone clock transition. For the $m_J=0\rightarrow m_{J'}=0$ transition, we obtain $C_2 =$\SI{0.39\pm0.04}{\hertz\per\milli\tesla\squared}, which is to our knowledge the smallest reported for any atomic transition to date. This confirms the predicted low sensitivity of highly charged ions to higher-order Zeeman effects, making them ideal candidates for high-precision optical clocks.
        Comparison of the experimental results with our state-of-the art electronic structure calculations shows good agreement, and demonstrates the significance of the frequency-dependent Breit contribution, negative energy states and QED effects on magnetic moments. 
	\end{abstract}
	
	\maketitle
	%TC:endignore    

	\textit{Introduction}—Highly charged ions (HCI) have extreme electronic properties as a result of strong internal electric fields, allowing precise tests of fundamental physics \cite{safronova_search_2018, kozlov_hci}. 
    Measurements of atomic parameters of these few-electron HCI are of interest, because theory predictions can reach accuracies far beyond what is possible in many-electron systems. In addition, quantum electrodynamics (QED) effects are greatly enhanced due to the high charge state, allowing stringent tests of QED in the strong-field regime \cite{blaum_perspectives_2020}. 
    Furthermore, the response of the atomic structure to a magnetic field $B$ can be both calculated and measured with high accuracy \cite{morgner_stringent_2023, sailer_direct_2022, heise_high-precision_2023, wagner_g_2013}. 
    
%\Section method
% \section result
    %The Zeeman effect is the energy shift of an electronic state to an external magnetic field $B$. 
    For an electronic level with total angular momentum $J$ and without hyperfine structure, the Zeeman shift from an external magnetic field $B$ is \cite{von_lindenfels_experimental_2013}
    \begin{equation}
        \Delta E_{m_J} = m_J g \mu_{\rm B} B + g^{(2)}(m_J) \frac{\left(\mu_{\rm B} B\right)^2}{m_ec^2} + \mathcal{O}(B^3) \, ,
        \label{eq:zeeman}
    \end{equation}
    with the magnetic quantum number $m_J$ and the $g$-factor of the state, the Bohr magneton $\mu_{\rm B} = e \hbar/2 m_e$, the Planck constant $h$, the electron rest mass $m_e$, the elementary charge $e$, and the speed of light $c$. The second-order Zeeman coefficient $g^{(2)}$ is $m_J$-dependent with the symmetry relation $g^{(2)}(-m_J)=g^{(2)}(m_J)$. 

    While investigations of the ground-state $g$-factor in HCI in Penning traps currently enable the most precise tests of strong-field QED \cite{morgner_stringent_2023}, these types of measurements cannot be easily transferred to excited states, since the measurement sequence typically takes longer than the excited state lifetime. 
    Measurements of excited-state $g$-factors in electron beam ion traps (EBIT) have relative uncertainties of the order of $10^{-4}$ \cite{rehbehn_sensitivity_2021}, and thus larger than those of theory predictions \cite{soria_orts_zeeman_2007}.
    Our recently developed HCI-based optical clock \cite{king_optical_2022} has enabled measurements of frequencies with sub-Hz precision, and of excited-state $g$-factors with $10^{-6}$ uncertainty \cite{micke_coherent_2020, king_optical_2022}. 
    However, the magnetic fields $B$ in those measurements were calibrated relative to the known ground state $g$-factor of \HCI{Ar}{13} \cite{arapoglou_g-factor_2019}.    Knowledge of the Zeeman shifts is also required to obtain unperturbed clock transition frequencies \cite{king_optical_2022, kozlov_hci}. 
    For clock transitions, the first-order linear Zeeman shift can be averaged to zero \cite{bernard_laser_1998} by measuring components with opposite $m_J$.
    The quadratic terms in the second-order Zeeman shift preclude such procedure, and thus one has to correct for the differential shift $\Delta\nu=C_2(m_J, m_J') B^2$ between the excited ($m_J'$) and ground state ($m_J$)  \cite{itano_external-field_2000} with
    \begin{equation}
        C_2(m_J, m_{J'}) = \frac{\mu_{\rm B}^2}{m_ec^2h}(g^{(2)}_\mathrm{e}\left(m_{J'}\right)- g^{(2)}_\mathrm{g}\left(m_{J}\right)).
        \label{eq:C2}
    \end{equation} 
    %in units of \SI{}{\hertz\per\tesla\squared}. 
    For HCI without hyperfine-structure, the sparsity of low-lying energy states due to the large fine-structure splitting should lead to much smaller $C_2$ than in neutral or singly charged systems \cite{berengut_highly_2012}, but experimental confirmation was so far lacking. 

\begin{figure*}[tb!]
		\centering
		\includegraphics[width=0.8\textwidth]{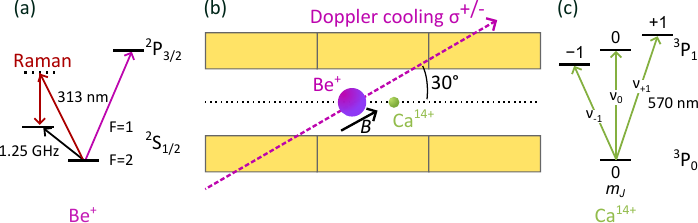}
		\caption{Sketch of the experimental setup with simplified level schemes of \Be and \HCI{Ca}{14}: (a) \Be with Doppler cooling transition $^2\!S_{1/2} \rightarrow \,^2\!P_{3/2}$ at \SI{313}{\nm} (purple); hyperfine ground state $^2S_{1/2}$ transition $F=2\rightarrow F=1$ at \SI{1.25}{\giga\hertz} is driven either by microwaves (black) or a stimulated Raman process (red). (b) A \Be-\Ca\ two-ion crystal confined in a linear Paul trap. The quantization axis at an angle of ca. $30^\circ$ relative to the symmetry axis of the trap is defined by $B$ as set by three orthogonal pairs of coils (not shown) and optional NdFeB permanent magnets (not shown). The propagation of the Doppler-cooling laser is parallel to $B$. 
        (c) Simplified level scheme of \HCI{Ca}{14}; $\nu_{-1,0,+1}$ label the Zeeman components of the \threePzero, $m_J =0$ to \threePone, $m_J=-1,0,1$ clock transition at a wavelength of \SI{570}{\nano\meter}. }
		\label{fig:setup}
	\end{figure*}

    In this Letter, we report on measurements and atomic structure calculations of the excited-state \threePone $g$-factor and $C_2(0,0)$ of the \threePzero-\threePone transition in \Ca. Following the experimental approach of Ref.~\cite{rosenband_observation_2007}, the knowledge of the magnetic field is derived from a co-trapped \Be ion. We achieve a relative uncertainty of the \HCI{Ca}{14} $g$-factor of $4\times10^{-6}$, which, by comparison to calculations, resolves the QED and negative energy eigenstate contributions to $g$ in a system with as many as six electrons. For the calculations, we demonstrate the convergence of the configuration interaction computation in this six-electron system, which enabled us to show the significance of the frequency-dependent Breit contribution to predicting energies of optical transitions in HCI. The small, measured $C_2(0,0)$ confirms the predicted low sensitivity of HCI to higher-order magnetic field effects. The experimental results provide a testbed for theoretically predicted excited-state magnetic field properties that can easily be transferred to other HCI with optical transitions for the development of high-accuracy HCI clocks. 

    \textit{Experimental setup}— A \Be\ and a \Ca\ ion are confined together in a cryogenic linear Paul trap \cite{leopold_cryogenic_2019, micke_closed-cycle_2019} as sketched in Fig.~\ref{fig:setup}. The \Be ion is Doppler cooled using its $\mathrm{S}_{1/2}\rightarrow \mathrm{P}_{3/2}$ transition at \SI{313}{\nano\meter}. The hyperfine transition at \SI{1.25}{\giga\hertz} between $\mathrm{S}_{1/2}, F=2$ and $\mathrm{S}_{1/2}, F=1$ is driven either by a microwave (mw) antenna close to the ion trap or by an optically stimulated Raman transition addressing motional sidebands. The frequencies of the hyperfine transitions between the magnetic field-sensitive states $F=2, m_F=\pm2$ and $F=1, m_F=\pm1$ are used throughout this work to calibrate the magnetic field at the \Be\ position with high accuracy. For this, we numerically invert the Breit-Rabi formula and use the accurately known \Be $g$-factors (for more details about this procedure, see \cite{supplementary}) \cite{wineland_laser-fluorescence_1983, shiga_diamagnetic_2011}. All used radio and mw frequencies are referenced to a calibrated H-maser of PTB. 
	
	Production, recapture, and cooling of HCI are described in our previous work \cite{micke_heidelberg_2018, wilzewski_nonlinear_2024, schmoger_coulomb_2015}. In brief, a single \Be ion co-trapped with an HCI provides sympathetic cooling and enables quantum logic spectroscopy of the latter with sub-Hz precision \cite{micke_coherent_2020, schmidt_spectroscopy_2005}.
    Here, we study $^{40}$\Ca with an optical transition \threePzero $\rightarrow$ \threePone\ at \SI{570}{\nano\meter} that features an excited-state lifetime of \SI{11}{\milli\second}. In an external magnetic field, its three Zeeman components \threePzero, $m_J =0$ $\rightarrow$ \threePone, $m_{J'}=0,\pm 1$ have $\nu_{0,\pm 1}$ transition frequencies (see Fig.~\ref{fig:setup}(c)). These transitions are interrogated with a laser that is stabilized \cite{stenger_ultraprecise_2002} to an ultrastable Si2 cavity \cite{matei_1.5_2017} with an optical frequency comb. 
    	
	The quantization axis is defined at an angle of approximately \SI{30}{\degree} relative to the symmetry axis of the trap by a magnetic field $B$, as shown in Fig.~\ref{fig:setup}(b). The direction of $B$ is aligned to a fixed $\sigma^{+/-}$-polarized laser beam that performs Doppler cooling and state detection of the \Be ion. Three orthogonal pairs of coils set the magnetic field direction and strength between \SI{-160}{\micro\tesla} and \SI{350}{\micro\tesla}. Here, the sign of the $B-$field indicates its direction relative to that of the laser propagation. For stronger fields up to \SI{1.7}{\milli\tesla}, a pair of NdFeB permanent magnets are placed symmetrically around the exit port of the Doppler-cooling beam. %This is limited by the size of the vacuum chamber (radius $\approx$ \SI{30}{\centi\meter}). 
    The static magnetic field is actively stabilized using a commercial fluxgate magnetometer outside the vacuum chamber and an additional set of compensation coils \cite{leopold_cryogenic_2019}, reaching a fractional instability of $10^{-5}$ for up to \SI{100}{s} at the ion position.
    
    \begin{figure*}
		\centering
		\includegraphics[width=\textwidth]{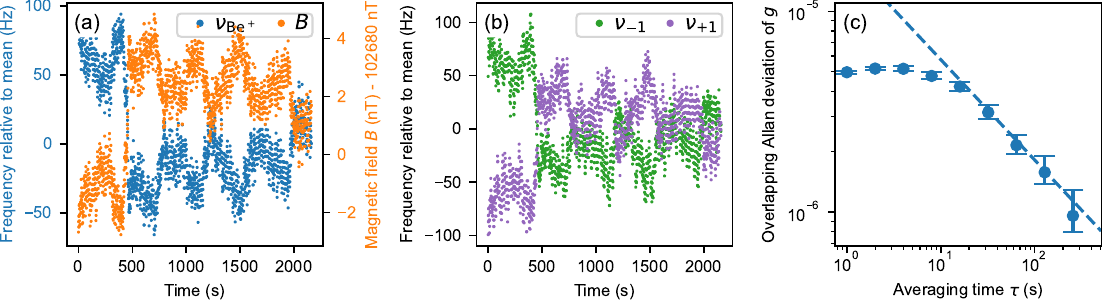}
		\caption{(a): Frequency deviation of the hyperfine transition $\nu_{\mathrm{Be}^+}$ in \Be\ (blue) and the derived magnetic field $B_{\mathrm{Be}}$ (orange). Oscillations at a period of \SI{6}{\minute} are caused by variations of the temperature in the laboratory. (b) Frequency deviations of the magnetic field-sensitive transitions $\nu_{-1}$ (green) and $\nu_{+1}$ (purple) in \Ca.  From the data shown in (a) and (b), the $g$-factor is derived using Eq. \eqref{eq:g_factor}. %Bottom: Result of the two independent $g$-factor measurements.
        (c) Relative measurement instability of the $g$-factor quantified by the overlapping Allan deviation. The dashed line is a fit $\propto \tau^{-1/2}$, as expected from quantum projection noise for long averaging times.}
        \label{fig:g_factor}
	\end{figure*}
 
	\textit{$g$-factor of the \threePone-state}—The $g$-factor of the \threePone\ state is derived from the first-order Zeeman shift in \Ca through monitoring of $\nu_{\pm1}$ while measuring $B$ through the hyperfine transition in \Be. The magnetic field at the position of the \HCI{Ca}{14} ion is obtained from that at the \Be position $B_{\mathrm{Be}}$ \cite{shiga_diamagnetic_2011} and a correction for its gradient along the trap axis $b_z=\mathrm{d}B/\mathrm{d}z$ with
    \begin{equation}
        g = \frac{h(\nu_{+1}-\nu_{-1})}{2\mu_\mathrm{B}(B_{\mathrm{Be}}+b_z d)}.
        \label{eq:g_factor}
    \end{equation}
    The \Ca-\Be distance $d$ is derived from measurements of the axial motional frequencies of the \Be-\Ca\ crystal (for details about the measurements and calculations of $d$, see Supplemental Material \cite{supplementary}) \cite{micke_coherent_2020, leopold_cryogenic_2019,heinzen_quantum-limited_1990, mccormick_coherently_2019, wunderlich_robust_2007, james_quantum_1998, kielpinski_sympathetic_2000,home_normal_2011}. 
    %The gradient ($b_z=\mathrm{d}B/\mathrm{d}z$) along the \Be-\Ca\ direction is measured in an independent measurement described below. 
    
    The field gradient $b_z=$~\SI{0.30\pm0.02}{\nano\tesla\per\micro\meter} was determined by moving a \Be ion along the axial symmetry axis of the trap and probing the mw transition to determine the magnetic field. 
    We also probe at different radial positions, because varying transverse electric fields lead to charge-to-mass ratio-dependent radial displacements, tilting the \Ca-\Be crystal axis with respect to the symmetry axis of the trap \cite{barrett_sympathetic_2003}, which is the dominant uncertainty of $b_z$. Further details are given in the Supplemental Material \cite{supplementary}.
    All parameters of Eq.~\eqref{eq:g_factor}, except for $b_z$, are measured in parallel. 
    
    Data obtained under typical operating conditions of $B_{\mathrm{Be}}\approx$~\SI{103}{\micro\tesla} and $d\approx$~\SI{20.4}{\micro\meter} is shown in Fig.~\ref{fig:g_factor}.
    In Fig.~\ref{fig:g_factor}(a), the deviations of the \Be hyperfine transition frequency from its mean and the derived magnetic field are shown. 
    Fig.~\ref{fig:g_factor}(b) shows the recorded frequency deviation of $\nu_+$ and $\nu_-$ in \Ca in relation to their mean frequency. Additionally, simultaneous measurements of the motional frequency yielded a time-resolved value for $d$ (see Supplemental Material \cite{supplementary}). %With this data, the $g$-factor of the \threePone state is derived using Eq.~\eqref{eq:g_factor}. 
    The overlapping Allan deviation of the $g$-factor is shown in Fig.~\ref{fig:g_factor}(c), showing an instability as expected from white-frequency noise for long averaging times. We assume such a behavior in the entire dataset for determining its statistical uncertainty.
    
    Two independent runs yield a mean value of $g=1.499032(6)$ with a relative uncertainty of $4\times10^{-6}$. The error budget is shown in Table~\ref{tab:g_uncert}. The largest contribution arises from the magnetic field gradient. 
    Installation of gradient compensation coils could reduce the uncertainty to the low $10^{-7}$ level in the future, where the accuracy of the ground-state hyperfine structure of \Be will become the limiting factor. 
    We have measured the trap drive-induced a.c. magnetic field \cite{trap_ac_zeeman, gan_oscillating-magnetic-field_2018, arnold_precision_2020} and determined its influence on the $g$-factor measurement to be negligible.

	\textit{$C_2$ coefficient}—For the measurement of % the second-order Zeeman coefficient
    $C_2(0,0)$ in \HCI{Ca}{14}, we employ the first-order magnetic field-insensitive transition \threePzero, $m_J=0 \rightarrow \threePone, m_{J'}=0$ with the transition frequency $\nu_0$. The frequency shift of $\nu_0$ is $\Delta \nu_0 = C_2(0,0)B^2$.
    We measured $\nu_0$ in four magnetic fields from \SI{25}{\micro\tesla} to \SI{1.65}{\milli\tesla} using a Yb$^+$ single-ion optical clock as a reference \cite{huntemann_single-ion_2016, king_optical_2022}. For the two largest magnetic field measurements, we employed additional NdFeB magnets attached to one side of the vacuum chamber. The measurements of $\nu_0$ were performed for at least \SI{15000}{\second}, and reached a statistical uncertainty of approximately \SI{100}{\milli\hertz} for each magnetic field setting.
    The magnetic field strength $B$ is again obtained from the hyperfine transition frequency measurement in \Be. We have confirmed experimentally that, despite the larger magnetic field gradient here, $B_\mathrm{Ca}$ agrees with $B_\mathrm{Be}$ to better than $1\%$ (see Supplemental Material for further experimental details \cite{supplementary}). Thus, the corresponding gradient correction is negligible in comparison to the statistical uncertainty of the frequency measurements. 
    
\begin{figure}[!t]
		\centering
		\includegraphics[width=\columnwidth]{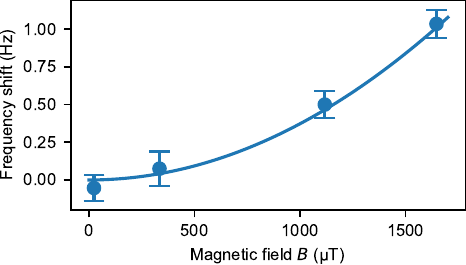}
		\caption{Measurement of the $m_J=0\rightarrow m_{J'}=0$ clock transition frequency shift in \Ca\ as a function of magnetic field strength $B$. The fit of the form $\Delta\nu_0=C_2(0,0)B^2$ yields $C_2(0,0)=$ \SI{0.39\pm0.04}{\hertz\per\milli\tesla\squared}. }
		\label{fig:c2ca14}
	\end{figure}

	The frequency shift $\Delta\nu_0$ of the \HCI{Ca}{14} clock transition at different magnetic field strengths is shown in Fig.~\ref{fig:c2ca14}. A quadratic fit of the form $\Delta\nu_0=C_2(0,0)B^2$ yields $C_2(0,0)=$ \SI{0.39\pm0.04}{\hertz\per\milli\tesla\squared}, where the uncertainty is derived from the fit. The systematic uncertainty from shifts of the \HCI{Ca}{14} clock transition, the reference clock \cite{huntemann_single-ion_2016}, the frequency comb and other parts of the frequency chain are negligible. 
    
    Assuming our typical d.c.\!~magnetic field of \SI{25}{\micro\tesla} for \Ca clock operation \cite{wilzewski_nonlinear_2024}, the measured $C_2(0,0)$ yields a fractional shift of $4.6(5)\times10^{-19}$.

    \begin{table}
        \centering
        \caption{Summary of $g$-factor measurement uncertainties\label{tab:g_uncert}}
        \begin{tabularx}{\columnwidth}{ Xc } 
        \hline\hline
         Source & Uncertainty / $10^{-6}$ \\
         \hline
         Gradient & $6$  \\ 
         Statistics & $0.4$ \\
          \Be\ atomic parameters & $0.2$  \\ 
         Trap drive-induced a.c. Zeeman & $<0.1$  \\
         Ion-ion distance & $<0.1$ \\
         \hline\hline
        \end{tabularx}
    \end{table}
    
    \textit{Theory}—
    The measured atomic parameters are compared with atomic structure calculations. This requires the calculation of transition energies as an essential quality test of the wave functions used to compute the $g$-factors.
    The calculations 
    %of the energies and $g$ factors 
    are performed using a large-scale configuration interaction (CI) method to correlate the six electrons following Refs.~\cite{sym2021,Fe2024,2025pCI}. We converge the CI computation, including excitations up to $24spdf\!ghi$ and extrapolating contributions of higher partial waves.
    QED corrections are calculated according to Ref.~\cite{QED}. The results of the computations are listed in Table~\ref{energies}. We find an unexpectedly large contribution of the frequency-dependent Breit interaction corrections, which is listed separately in the table. 
     
    Our calculations show that frequency-dependent Breit contributes at the level of 1\% to the Breit interaction, enabling us to estimate for which cases frequency-dependent Breit contributions should be computed to achieve the expected accuracy in future HCI computations. 
     The resulting theoretical energy values are in excellent agreement with the experimental values, as shown in Table~\ref{energies}. The details of the calculations are described in the Supplemental Material \cite{supplementary}. 
	
    The calculated $g$-factors of the low-lying states in \Ca are listed in the last column of Table~\ref{energies}. 
    To calculate them, we use the operator for the interaction with the external homogeneous magnetic field $\bm{B}$, which is assumed to be aligned with the $z$-axis:
    \begin{equation}
    \label{magnetic_interaction_operator}
    % V_m = {\bm \mu}^{(1)} \cdot {\bm B} = {\mu}^{(1)}_z  {B}_z\, ,
    V_m = {\mu}_z  {B}_z = \frac{ec}{2} \,  \sum_{i} (\bm{r}_i \times \bm{\alpha}_i)_z B_z \, ,
    \end{equation}
    %
    %
    % where 
    % %
    % %
    % \begin{equation}
    % \label{eq:mu_operator}
    % {\bm \mu}^{(1)} = \frac{ec}{2} \,  \sum_{i} (\bm{r}_i \times \bm{\alpha}_i) \,
    % \end{equation}
    %
    %
    where $\mu_z$ is the atomic magnetic moment and $\bfalpha_i$ is the vector of the Dirac matrices for the $i^{th}$ electron of an ion.
    \begin{table}
    \caption{\label{energies} Theoretical energies (in cm$^{-1}$) and $g$-factors of $2s^2 2p^2$ states of Ca$^{14+}$ given relative to the $^3$P$_0$ ground state energy. 
   The contributions of frequency-dependent Breit and QED are given in columns `freq.' and `QED', respectively.  
    The differences between the final energies with the NIST and Ref.~\cite{PRA} data are given in columns, $\Delta^{a}$ and $\Delta^{b}$, respectively. $g$-factors are given in the last column. 
    }
    \begin{ruledtabular}
    \begin{tabular}{lccccccc}
    \multicolumn{1}{c}{Conf.}&          
    \multicolumn{1}{c}{CI}& 
    \multicolumn{1}{c}{freq.}&
    \multicolumn{1}{c}{QED}&
    \multicolumn{1}{c}{Final}& 
    \multicolumn{1}{c}{$\Delta^{a}$}& 
    \multicolumn{1}{c}{$\Delta^{b}$} &
    \multicolumn{1}{c}{$g$-factor}
    \\
    \hline \\[-0.7pc]
    $^3\text{P}_1$ &    17507     &  -10 & 60  &  17557  &    -2 &  1  &  1.49902  \\
    $^3\text{P}_2$ &    35839     &  -17 & 99  &  35921  &    -2 &  9  &  1.47075  \\
    $^1\text{D}_2$ &   108555     &  -18 & 111 & 108648  &    48 &     &  1.02575  \\
    $^1\text{S}_0$ &   197726     &  -21 &  52 & 197757  &    87 &     & 
      \end{tabular}
    \end{ruledtabular}
    \end{table}
    The operator in Eq.\eqref{magnetic_interaction_operator} mixes the large and small components of the wave functions. In this case, negative energy states also contribute~\cite{Lindroth1993}. 
    
    The inclusion of QED effects in the Hamiltonian affects the wave functions and is crucial to obtaining energies that agree with the experiment (cf. Tab.~\ref{energies}). However, this plays a minor role for the $g$-factors. A much larger contribution to the $g$-factors comes from the QED corrections to the atomic magnetic moment and can be approximately estimated as an expectation value of the operator
    \begin{equation}
        \label{eq:delta_mu_operator}
        \Delta \mu_z = \, \frac{g_{\rm free}-2}{2}\mu_{\rm B}\sum_i \beta_i \Sigma_{z,i}, 
    \end{equation}
    where $\beta$ is the Dirac matrix,
    % \begin{equation*}
    % \beta = 
    % \begin{pmatrix}
    %      I &  0 \\
    %      0 & -I 
    % \end{pmatrix}, \quad  \quad  
    $\Sigma_z =
    \begin{pmatrix}
         \sigma_z  & 0 \\
            0      & \sigma_z 
    \end{pmatrix}
    $,
    % \end{equation*} 
     $\sigma_z$ is the Pauli matrix and $g_{\rm free} = 2[1+0.5(\alpha/\pi)-0.328478\ldots\times(\alpha/\pi)^2+\ldots]$ is the free-electron $g$-factor. 
    A good agreement between this estimate and the rigorous QED calculation was recently demonstrated in boron-like Ar~\cite{Maison2019}.
    For \HCI{Ca}{14}, the theoretical result for the \threePone state $g^{\rm (theo)}=1.49902$ is in good agreement with the experimental value $g=1.499032(6)$. We emphasize that the contributions from negative energy eigenstates (-0.00009) and QED corrections to the atomic magnetic moment (0.00116) are critical to achieve agreement. Further calculations should include the rigorous QED treatment and nuclear recoil corrections.   
    
    The calculation of $C_2$ is directly related to the second-order Zeeman coefficients $g_g^{(2)}$ and $g_e^{(2)}$ of the ground- and excited-ionic sublevels, following Eq.~(\ref{eq:C2}). For a particular sublevel $| \Gamma J m_J \rangle$, these dimensionless coefficients can be obtained within second-order perturbation theory as:
    \begin{align}
    \begin{split}
        \label{eq:g2_coefficient}
        g^{(2)}(m_J) = \frac{m_e c^2}{\mu_B^2} \sum_{\Gamma', J'} & 
        \frac{\left|\mem{\Gamma' J' m_J}{{\mu}_z + \Delta {\mu}_z}{\Gamma J m_J}\right|^2}{E(\Gamma J) - E(\Gamma' J')}
    \end{split},
    \end{align}
    where the operators ${\mu}_z$ and $\Delta {\mu}_z$ are given by Eqs.~(\ref{magnetic_interaction_operator}) and (\ref{eq:delta_mu_operator}), respectively. $J$ is the total angular momentum, and $\Gamma$ represents the set of all other quantum numbers necessary for a unique specification of the state. Moreover, the intermediate state summation $\sum_{\Gamma', J'}$ runs over all states with energies $E(\Gamma' J') \neq E(\Gamma J)$. For computational purposes, this summation was restricted to energetically near states, and the matrix elements in Eq.~(\ref{eq:g2_coefficient}) were obtained using the multiconfiguration Dirac-Fock (MCDF) approach. Based on MCDF calculations, we obtained the theoretical prediction $C_2^{\rm (theo)}(0,0) = $~\SI{0.373\pm0.0003}{\hertz\per\milli\tesla\squared} \cite{gilles_quadratic_2024}, which is in good agreement with the experimental value $C_2(0,0)=$~\SI{0.39\pm0.04}{\hertz\per\milli\tesla\squared}. To our knowledge, this is the first experimental confirmation of a theoretically calculated $C_2$ in HCI.
    
	\textit{Conclusions}— In this work, we have investigated the response of the atomic structure of carbon-like \HCI{Ca}{14} to magnetic fields through microwave and optical spectroscopy. 
    We measured the $g$-factor of the excited state \threePone with a relative uncertainty of $4\times 10^{-6}$. The state-of-the-art calculations show a good agreement between experiment and theory, highlighting the significance of QED contributions in the $g$-factor of few-electron systems, which have been largely unexplored in experimental studies before.
    Furthermore, we determined the second-order Zeeman shift coefficient $C_2(0,0)$, which, to the best of our knowledge, is the smallest value reported to date. 
    This proves the long-standing theoretical prediction that HCI are highly insensitive to external magnetic fields \cite{berengut_highly_2012}, and advances the work towards the development of next-generation optical clocks employing HCI.
    The experimental methods employed in this work are transferable to a wide range of HCI with an optical transition suitable for quantum logic spectroscopy \cite{kozlov_hci}. This includes hydrogen-like systems in heavy HCI \cite{schiller_hydrogenlike_2007}, where the ground-state $g$-factor currently provides the most stringent test of strong field QED \cite{morgner_stringent_2023}.
 
     \begin{acknowledgments}
    \textit{Acknowledgment}— We thank Mikhail Kozlov for bringing our attention to the potentially significant effect of the frequency-dependent Breit and subsequent fruitful discussions. We thank Ilya Tupitsyn and Dmitry Glazov for fruitful discussions regarding frequency-dependent Breit contributions and $g$-factor calculations. We thank Vladimir Yerokhin for computing the effect of the frequency-dependent Breit interaction in Li-like Ca, which demonstrated that the effect may be significant. We thank Daniele Nicolodi, Thomas Legero and Uwe Sterr for providing the ultra-stable Si cavity as a laser reference.
    The project was supported by the Physikalisch-Technische Bundesanstalt, the Max-Planck Society, the Max-Planck–Riken–PTB–Center for Time, Constants and Fundamental Symmetries, and the Deutsche Forschungsgemeinschaft (DFG, German Research Foundation) through SCHM2678/5-2, SU 658/4-2, the collaborative research centers SFB 1225 ISOQUANT and SFB 1227 DQ-\textit{mat}, and under Germany’s Excellence Strategy – EXC-2123 QuantumFrontiers – 390837967. The project 20FUN01 TSCAC has received funding from the EMPIR programme co-financed by the Participating States and from the European Union’s Horizon 2020 research and innovation program. This project has received funding from the European Research Council (ERC) under the European Union’s Horizon 2020 research and innovation program (grant agreement No 101019987).
    The theoretical work has been supported in part by the US NSF Grant  No. PHY-2309254,  US Office of Naval Research Grant No. N00014-20-1-2513, and by the European Research Council (ERC) under the Horizon 2020 Research and Innovation Program of the European Union (Grant Agreement No. 856415).
    The calculations in this work were done through the use of Information Technologies resources at the University of Delaware, specifically the high-performance Caviness and DARWIN computer clusters.     
 \end{acknowledgments}

	%\bibliographystyle{apsrev4-2} % Tell bibtex which bibliography style to use
	%\bibliography{Bibliography.bib,theory} 
%\input{main.bbl}

\end{document}

\title{Supplemental material for Excited-state magnetic properties of carbon-like \Ca}	
   \author{Lukas J. Spieß}
	\author{Shuying Chen}
    \email{shuying.chen@quantummetrology.de}
	\author{Alexander Wilzewski}
	\author{Malte Wehrheim}
	\affiliation{Physikalisch-Technische Bundesanstalt, Bundesallee 100, 38116 Braunschweig, Germany}

    \author{Jan Gilles}
	\author{Andrey Surzhykov}
	\affiliation{Physikalisch-Technische Bundesanstalt, Bundesallee 100, 38116 Braunschweig, Germany}
	\affiliation{Technische Universität Braunschweig, Universitätsplatz 2, 38106 Braunschweig, Germany}
	
	\author{Erik Benkler}
	\author{Melina Filzinger}
 	\author{Martin Steinel}
	\author{Nils Huntemann}
	\affiliation{Physikalisch-Technische Bundesanstalt, Bundesallee 100, 38116 Braunschweig, Germany}
 
    \author{Charles Cheung}
    \affiliation{Department of Physics and Astronomy, University of Delaware, Newark, Delaware 19716, USA}
    
    \author{Sergey G. Porsev}
    \affiliation{Department of Physics and Astronomy, University of Delaware, Newark, Delaware 19716, USA}
    
    \author{Andrey I. Bondarev}
    \affiliation{Helmholtz-Institut Jena, 07743 Jena, Germany}
    \affiliation{GSI Helmholtzzentrum f\"ur Schwerionenforschung GmbH, 64291 Darmstadt, Germany}
    
    \author{Marianna S. Safronova}
    \affiliation{Department of Physics and Astronomy, University of Delaware, Newark, Delaware 19716, USA}

    \author{José R. Crespo López-Urrutia}
	\affiliation{Max-Planck-Institut für Kernphysik, Saupfercheckweg 1, 69117 Heidelberg, Germany}
	
	\author{Piet O. Schmidt}
	\email{piet.schmidt@quantummetrology.de}
	\affiliation{Physikalisch-Technische Bundesanstalt, Bundesallee 100, 38116 Braunschweig, Germany}
	\affiliation{Institut für Quantenoptik, Leibniz Universität Hannover, Welfengarten 1, 30167 Hannover, Germany}
	
	\begin{abstract}
		In this document, we provide supplementary information on the measurements and calculations performed.
	\end{abstract}
	
	\maketitle

    \section{Breit-Rabi formula}

    The Breit-Rabi formula is applicable to states with $J=\frac{1}{2}$ and nuclear spin $I$ in an external magnetic field $B$. The energy of the state $E(F,m_F)$ is then given by \cite{shiga_diamagnetic_2011}
    \begin{align}
        E\left(I\pm \frac{1}{2},m_F\right)=&hA\bigg(-\frac{1}{4} + \frac{g_I m_F \mu_\mathrm{B} B}{hA} \\
        \pm& \frac{2I+1}{4}\sqrt{1+\frac{4m_F}{2I+1}X+X^2}\bigg)\nonumber
    \end{align}
    where $X=\mu_\mathrm{B}B(g_J-g^\prime_I)/\left[(I+1/2)hA\right]$ and $\mu_\mathrm{B}$ is the Bohr magneton. The specific constants for \Be are the hyperfine constant $A=\,$\SI{-625008837.044(12)}{\hertz}, the nuclear to electronic $g$-factor ratio $g^\prime_I/g_J=\,$\SI{2.134 779 852 7(10)e-4}{} and the electronic $g$-factor $g=\,$\SI{2.00226239(31)}{}, taken from \cite{shiga_diamagnetic_2011}.
    
    \section{HCI injection at high magnetic field}

    For an increasing magnetic field, we observe a decreasing HCI loading efficiency when using permanent magnets. For magnetic field strengths above \SI{1.5}{\milli\tesla} at the trapping center, we were unable to retrap HCI. The effect was more pronounced when the magnets were placed close to the beamline. We attribute this to a change in the HCI trajectory due to a Lorentz force. Measurements above \SI{1.5}{\milli\tesla} required loading of an HCI first, before the magnets could be installed.
        
    \section{Measurement of the motional frequency}

\begin{figure}
		\centering
		\includegraphics[width=\columnwidth]{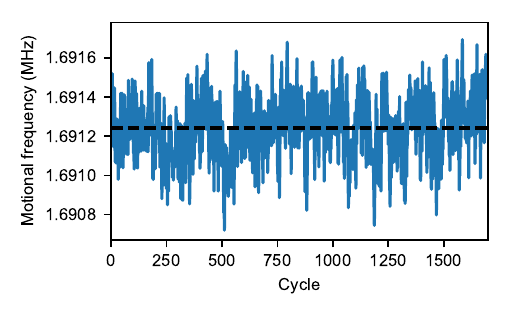}
		\caption{Measurement of the motional frequency of a \Ca-\Be\ crystal. The black dashed line corresponds to mean value of the motional frequency. }
		\label{fig:motional_freq_trace}
	\end{figure}

    To measure the motional frequency of any given mode, the ion crystal is initially ground-state cooled using resolved-sideband cooling \citep{micke_coherent_2020, leopold_cryogenic_2019}. Then, an oscillating electric field applied to a spare electrode displaces the motional state, with an amplitude that depends on the detuning of the applied electric field from the motional frequency \cite{heinzen_quantum-limited_1990, mccormick_coherently_2019}. The displaced state occupies several Fock states with varying Rabi frequency. To achieve an optimal mapping of the motional excitation to the electronic state, we employ a rapid adiabatic passage \cite{wunderlich_robust_2007} on the first-order red sideband on the Raman transition of \Be. Rapid adiabatic passage has, compared to Rabi excitation, the advantage that it is largely insensitive to variations of the Rabi frequency. A typical time-resolved motional frequency measurement trace is shown in Fig.\ref{fig:motional_freq_trace}.

    \section{Ion-ion distance calculation} \label{sec:ioniond}
    
    In a linear Paul trap the ion-ion distance $d$ of an axially oriented crystal can be calculated from a minimization of the potential energy \cite{james_quantum_1998, kielpinski_sympathetic_2000}. Assuming a general mixed two-ion crystal in a purely harmonic potential, this yields:
    \begin{equation}
        d = \left(1+\beta^{-1}\right)\left(\frac{q_1 q_2}{4\pi\epsilon_0\left(1+\beta^{-1}\right)^2 m_1 \omega_{z,1}^2}\right)^{1/3}, \label{eq:distance}
    \end{equation}
    where $q_i$ is the charge of ion $i=1,2$, $m_i$ is the ion mass, $\epsilon_0$ is the dielectric constant and $\omega_{z,i}$ is the axial motional frequency when only a single ion of type $i$ is trapped. We also introduced the charge ratio $\beta=q_2/q_1$.
    
    The single-ion motional frequency $\omega_{z,i}$ can be measured directly with motional spectroscopy of a single trapped \Be, or it can be calculated from a measurement of the coupled motional mode frequency \cite{barrett_sympathetic_2003}. The latter is employed during our measurements involving two-ion crystals to get an estimate of $d$. The out-of-phase motional frequency $w_{z,\mathrm{OOP}}$ of the two-ion crystal is related to $\omega_{z,i}$ by
    \begin{equation}
        w_{z,\mathrm{OOP}} = \omega_{z,1} \sqrt{\frac{\rho+\kappa}{2\left(\beta+1\right) \mu}},
    \end{equation}
    where we introduced the mass ratio $\mu=m_2/m_1$ and the parameters
    \begin{align}
        \rho &= \beta^2 + 3\beta\left(\mu+1\right) + \mu, \\
        \kappa &= \sqrt{\rho^2-12\mu\beta\left(\beta+1\right)^2}.
    \end{align}
    For our typical trapping parameters, the \Be-\Ca\ distance is $d\approx$~\SI{20.4}{\micro\meter}.% This is much larger than singly-charged ions due to the high charge of \Ca.

    In the experiment, the ion-ion distance derived using Eq. \eqref{eq:distance} deviates from the real value due to two effects. Firstly, a transverse electric field leads to a tilting of the crystal, which modifies the motional frequencies. We can observe this experimentally by deliberately applying a transverse electric field, where the maximum or minimum of the motional frequency indicates alignment of the crystal with the trapping axis. Our usual operating point of micromotion compensation corresponds to a deviation in motional frequency of less than \SI{1}{\kilo\hertz} from its optimal point. This is taken as a full uncertainty, yielding a \Be-\Ca\ distance uncertainty of \SI{8}{\nano\meter}. 
    
    Secondly, the anharmonic terms of the trapping potential change the motional frequencies and the ion-ion distance \cite{home_normal_2011}. To estimate the magnitude of the anharmonicities, we compare the frequency of the center-of-mass mode of ion chains composed of one, two and three \Be\ ions \cite{home_normal_2011}, yielding a small third-order contribution to the trapping potential $\propto z^3$. The \Be-\Ca\ crystal's motion is simulated by numerically integrating the equations of motion with and without an anharmonic term. From this we can bound the shift of $d$ to $<$~\SI{1}{\nano\meter}.
    
    \section{Magnetic field gradient measurements}

    The magnetic field gradient $b_z$ is derived from $b_z = \Delta B/d$, where $\Delta B$ is the magnetic field difference between two points in the trap separated by distance $d$. %The magnetic field is measured from probing the magnetic sensitive transition of the ion and $d$ is derived from the motional frequency measurement for single ion or two ion system, as described above.
    We used two different methods to measure the magnetic field gradient $b_z$ across the trap, either using a \Be-\Be\ two-ion crystal or sampling with a single \Be\ ion. Each method has advantages and disadvantages, which will be briefly discussed here.
    
    \subsection{Method 1: Using a \Be-\Be\ two-ion crystal}

\begin{figure}
		\centering
		\includegraphics[width=\columnwidth]{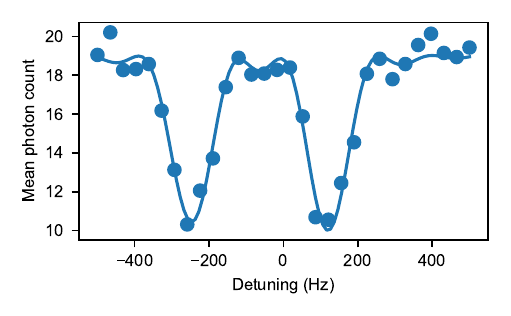}
		\caption{Measurement of the microwave transition of a \Be-\Be\ crystal. The frequency difference of \SI{368}{\hertz} corresponds to a magnetic field difference of \SI{17.5}{\nano\tesla}. The fit is given by two Rabi lineshapes.}
		\label{fig:twobemw}
	\end{figure}

    Trapping two \Be\ simultaneously enables a direct measurement of the magnetic field gradient along the \Be-\Be-axis by comparing their qubit transition frequency driven by the microwave antenna. The prerequisite is that the difference in the transition frequency is much larger than the FWHM linewidth $\delta\nu\approx0.8\Omega/\pi$ of the probed transition, where $\delta\nu$ is in Hertz and $\Omega$ is the on-resonant Rabi frequency given in \SI{}{\radian\per\second}. We assume Rabi excitation with an ideal $\pi$ rotation. An example of such a measured spectrum is shown in Fig. \ref{fig:twobemw}. 
    %The overlap of the the two transitions leads to a shift of the measured frequency difference which we investigate numerically and is found to be negligible for our purposes.
    Simultaneously to the transition frequency difference, we measure the center-of-mass motional frequency of the two-ion crystal,  which is used to infer the ion-ion distance (see above). 
    
    The method can be further improved by employing three ions, which gives access to the variation of the gradient of the magnetic field. Such measurements were performed at a gradient of \SI{2}{\nano\tesla\per\micro\meter} ($B\approx$ \SI{350}{\micro\tesla}) and the result is consistent with no change in the gradient with an upper bound on the absolute value of \SI{0.16e-3}{\nano\tesla\per\micro\meter\squared} in a confidence interval of $95\%$, which is negligible for the presented measurements.

    The advantage of using the \Be-\Be two-ion crystal method is that it is easily implemented and allows fast and highly precise measurements. Long-term measurements using this method show that the gradient is highly stable and does not vary significantly on time scales of several days at the subpercentage level. The downside of this method is that the gradient is measured along the \Be-\Be\ axis, which does not necessarily align with the \Be-\Ca axis. In particular, transverse electric fields induce differences between these two and lead to shifts that are difficult to assess (see the section “$g$-factor at high magnetic field” below).

    \subsection{Method 2: Sampling with a single \Be\ ion}

    Method 1 can only be applied for sufficiently large gradients and sufficiently low Rabi frequencies. The latter is predominantly limited by our short-term magnetic field stability, which limits the probe time to \SI{10}{\milli\second}, corresponding to $\delta\nu\approx$ \SI{80}{\hertz}, in the best case. Thus, for small gradients, the two peaks cannot be reliably distinguished. In this case, we move a single \Be\ in the trap and measure the difference in qubit frequency at varying positions to estimate the gradient.

    This is implemented by setting \Be at a starting position where the qubit transition frequency is measured. Then it is moved along the symmetry axis of the trap by roughly \SI{21}{\micro\meter} using the dc voltages applied to our Paul trap. The distance is derived from calibrated CCD camera images. Furthermore, we displace \Be in the transverse direction with the electrodes that are also used for compensation of excess micromotion. There, the qubit transition frequency is measured again. From the difference in magnetic field, the magnetic field gradient is estimated.

    The uncertainty of this method is predominantly limited by the stability of the short-term magnetic field of \SI{200}{\pico\tesla} at the time interval of our measurement of \SI{30}{\second}. Variations of the transverse electric field over time, e.g. from charging and discharging surfaces during HCI loading, require adjustments of our micromotion compensation voltages of typically around \SI{2}{V} corresponding to a displacement of \SI{2}{\micro\meter}. Thus, we move \Be in the transverse direction by at least \SI{4}{V}, corresponding to at least \SI{4}{\micro\meter}, using the compensation voltage. All measurements are equally weighted and the uncertainty of $b_z$ was taken to be the standard deviation of all measurement points.
    
    \section{$g$-factor at high magnetic field}

    Increasing the magnetic field amplifies the Zeeman shift. Thus, if the accuracy of the frequency measurements remains constant, it allows for more precise measurements of the $g$-factor. In addition to the measurements in the main manuscript, we performed measurements at a higher magnetic field of \SI{350}{\micro\tesla} by employing permanent magnets. The use of permanent magnets leads to large magnetic field gradients on the order \SI{2}{\nano\tesla\per\micro\meter} at an angle of $30^\circ$ relative to the symmetry axis of the trap. This gradient leads to a relative correction of the $g$-factor of ca. $2\times10^{-4}$. Thus, an accurate measurement of the gradient along the \Be-\Ca axis is necessary, which we measured using Method 1 given above.

    However, the measured gradient along the \Be-\Be axis is not easily related to the \Be-\Ca axis, since it is perturbed by residual transverse electric fields which tilt the \Be-\Ca crystal. A tilt of the order of \SI{1}{\degree} relative to the trap symmetry axis leads in the worst case to a variation of the derived $g$-factor at the $1\times10^{-6}$ level. In particular, we observe that HCI loading leads to a slow ($\approx$ \SI{2}{\hour}) time-dependent variation of the measured $g$-factor at this level, indicating significant charging and discharging of dielectric surfaces. This is confirmed by micromotion measurements on the HCI, also showing slow variations after HCI loading.

\begin{figure}
		\centering
		\includegraphics[width=\columnwidth]{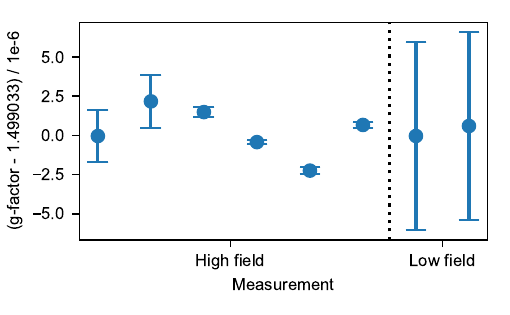}
		\caption{Measurement of the $g$-factor. The measurements at high magnetic field show systematic deviations which are attributed to a tilting of the ion crystal. This shift is not assessed here and is not included in the given error bars.}
		\label{fig:g_factor_all}
	\end{figure}

    The results of $g$-factor measurements at a higher magnetic field are shown in Fig. \ref{fig:g_factor_all}. We find good agreement between these measurements in a high magnetic field and the measurements in the main manuscript in the low magnetic field at the few $10^{-6}$-level, but scatter larger than the statistical and systematic uncertainty as a result of the crystal tilting, which was not taken into account. Nevertheless, the results from measurements at a high magnetic field give confidence in the results presented at low fields, although we refrain from assessing the uncertainty associated with the ion tilting. 
    In the low-field measurements, the effect of the tilting on the measurement is suppressed due to an almost seven times smaller gradient, which we account for by measuring $b_z$ using Method 2 given above.

    \section{Theory}

We start from the solution of the Hartree-Fock-Dirac equations in the central field approximation to construct the one-particle orbitals. The calculations are carried out using a configuration interaction (CI) method, which correlates all six electrons following \cite{Fe2024}. The CI many-electron wave function is obtained as a linear combination of all distinct states of a given angular momentum $J$ and parity:
$
\Psi_{J} = \sum_{i} c_{i} \Phi_i\,.
$
The energies and wave functions are determined from the time-independent many-electron Schr\"odinger equation 
$
H \Psi_n = E_n \Psi_n.
$
Breit interaction is included in all calculations. The QED corrections are calculated following Ref.~\cite{QED}.

 The main challenge of this computation is to ensure the numerical completeness of the CI expansion above, which requires large computational resources \cite{sym2021}.
 We use basis sets (i.e., the set of one-particle orbitals used in CI computations) of increasing size to check the numerical convergence of the values. 
  We label the basis set by the highest principal quantum number for each included partial wave. For example, $17spdf\!g$ means that all orbitals up to $n=17$ are included for the $spdf\!g$ partial waves. 
 The basis set is constructed in a radial cavity of 5 a.u. following recent studies where a compact basis set \cite{Fe2024} was shown to lead to significantly improved convergence with the principal quantum number $n$. 

The contributions to the energies of \Ca are listed in Table~\ref{energies}. The results are compared with the experimental data from the NIST database \cite{NIST} and from Ref.~\cite{PRA}. Note that the energies listed in column ``Ref.~\cite{PRA}'' are recomputed from experimental transition energies and converted from eV to cm$^{-1}$ using the recommended value of CODATA2018 of $hc$ \cite{CODATA2018}. It is essential to ensure that the energies are in good agreement with the experiment to investigate the quality of the wave function used to compute the $g$-factors. 

\begin{table*}
\caption{\label{energies} Contributions to the energies of Ca$^{14+}$. The results are compared with NIST and Ref.~\cite{PRA}. All energies are given in cm$^{-1}$. The basis set is designated by the highest quantum number for each included partial wave. For example, $17spdf\!g$ means that all orbitals up to $n=17$ are included for $spdf\!g$ partial waves. Contributions from extrapolation to higher ($ >\!k$) partial waves, triple excitations, frequency-dependent Breit, and QED are given separately. Contributions to the transition energies are given in the last two rows.}
\begin{ruledtabular}
\begin{tabular}{lccccccccccccccccc}
\multicolumn{2}{c}{Configuration}&
\multicolumn{1}{c}{NIST}&
\multicolumn{1}{c}{Ref~\cite{PRA}}&
\multicolumn{1}{c}{$17spdf\!g$}& 
\multicolumn{1}{c}{$+24g$}& 
\multicolumn{1}{c}{$+25h$}&
\multicolumn{1}{c}{$+24i$}&
\multicolumn{1}{c}{$+24k$}&
\multicolumn{1}{c}{$>\!k$}&
\multicolumn{1}{c}{triples}&
\multicolumn{1}{c}{extras}&
\multicolumn{1}{c}{freq. Br.}&
\multicolumn{1}{c}{QED}&
\multicolumn{1}{c}{Final}& 
\multicolumn{1}{c}{$\Delta^{a}$}& 
\multicolumn{1}{c}{$\Delta^{b}$}\\

\hline \\[-0.7pc]
$2s^2 2p^2$&$^3\!\text{P}_0$ &   0     &      0 &      0  &    0 &    0 &   0 &   0 &   0 &  0 &  0 &   0 &  0  &      0  &     0 &  0    \\
           &$^3\!P_1$ &   17559 &  17556 &   17500 &    2 &    5 &   2 &   0 &   0 & -1 & -1 & -10 & 60  &  17557  &    -2 &  1    \\
           &$^3\!P_2$ &   35923 &  35911 &   35839 &    2 &    0 &   0 &   0 &   0 & -1 & -1 & -17 & 99  &  35921  &    -2 &  9    \\
           &$^1\!D_2$ &  108600 &        &  108722 &   -4 &  -94 & -40 & -16 & -10 & -2 & -1 & -18 & 111 & 108648  &    48 &       \\
           &$^1\!S_0$ &  197670 &        &  197898 &  -14 & -126 & -55 & -22 & -14 & 27 & 32 & -21 &  52 & 197757  &    87 &       \\ [0.3pc]
\hline \\[-0.7pc]
$^3\!\text{P}_2 -\, ^3\!\text{P}_1$  &  &   18364 &  18355 &   18340 &    1 &   -5 &  -2 &  -1  & -1 &  0 &   0  &  -7 &  39 &  18364  &  0 &  8   \\
$^3\!\text{P}_1 -\, ^3\!\text{P}_0$  &  &   17559 &  17556 &   17500 &    2 &    5 &   2 &   1  &  0 & -1 &  -1  & -10 &  60 &  17557  & -2 &  1   \\
  \end{tabular}
\end{ruledtabular}
{\begin{flushleft}
\footnotesize
\tablenotemark{Difference with NIST}\\
\tablenotemark{Difference with Ref.~\cite{PRA}}\\
\end{flushleft}}
\end{table*}

\begin{table}
\caption{\label{energies-breit_comp} Contributions (in cm$^{-1}$) to the energies of Ca$^{14+}$ from the Breit effects. }
\begin{ruledtabular}
\begin{tabular}{llcccc}
\multicolumn{2}{c}{Configuration}&
\multicolumn{1}{c}{Coulomb}& 
\multicolumn{1}{c}{Breit}&
\multicolumn{1}{c}{freq. Br.}&
\multicolumn{1}{c}{freq. Br. \%}\\ 
\hline \\[-0.7pc]
$2s^2 2p^2$&$^3\!\text{P}_0$ &   0     &      0 &     0 &    0 \\
           &$^3\!\text{P}_1$ &   18323 &   -823 &   -10 & 1.2\% \\
           &$^3\!\text{P}_2$ &   38187 &  -2347 &   -17 & 0.7\% \\
           &$^1\!\text{D}_2$ &  110971 &  -2249 &   -18 & 0.8\% \\
           &$^1\!\text{S}_0$ &  199236 &  -1337 &   -21 & 1.6\% 
\end{tabular}
\end{ruledtabular}
\end{table}

\begin{table*}
\caption{\label{$g$-factors}  Contributions to the $g$-factors of Ca$^{14+}$ calculated in different approximations (see text). 
The basis set is designated by the highest quantum number for each included partial wave. For example, $24g$ means that all orbitals up to $n=24$ are included for the $spdf\!g$ partial waves. 
$g$-factors calculated from the nonrelativistic Land\'{e} formula are given in the column labeled
`NR' and relativistic values for a pointlike nucleus in the column labeled `Dirac'. The results are compared with the theoretical calculations of Ref.~\cite{Verdebout2014}.}
\begin{ruledtabular}
\begin{tabular}{llcccccccccc}
\multicolumn{2}{c}{Configuration}&
\multicolumn{1}{c}{NR}&
\multicolumn{1}{c}{Dirac}&
\multicolumn{1}{c}{$24g$}& 
\multicolumn{1}{c}{$+24h$}&
\multicolumn{1}{c}{$+24i$}&
\multicolumn{1}{c}{QED}&
\multicolumn{1}{c}{neg.}&
\multicolumn{1}{c}{QED$_\textrm{m}$}&  
\multicolumn{1}{c}{Final}&
 \multicolumn{1}{c}{Ref.~\cite{Verdebout2014}  }\\
 \hline \\[-0.7pc]
 $2s^2 2p^2$ &$^3\!\text{P}_1$ &1.5& 1.49734&  1.49795 & 0.00000 & 0.00000 &        0.00000 &-0.00009 &0.00116 & 1.49902  &  1.499146    \\
            &$^3\!P_2$ &1.5& 1.49734&  1.47006 & -0.00008 & -0.00003 & -0.00014 &-0.00015 &0.00109 & 1.47075  &  1.470979   \\
            &$^1\!D_2$ &1.0& 0.99679&  1.02562 & 0.00008 & 0.00003 &  0.00013 &-0.00017 &0.00006 & 1.02575  &  1.025764     
\end{tabular}
\end{ruledtabular}
\end{table*}

We start with all possible single and double excitations to any orbital up to $17spdf\!g$ from the $1s^2 2s^2 2p^2$ and $1s^2 2p^4$ configurations, correlating all six electrons. The results are listed in column
``$17spdf\!g$''. The basis set is then systematically expanded until a sufficient convergence of the energy levels is reached. For each partial wave $l$, contributions of orbitals with the highest principal quantum number $n$ are calculated until the differences are negligible. These contributions are listed under their respective ``$+nl$'' columns. 

Contributions from including additional four reference configurations, $2s 2p^2 3d$, $2s 2p^2 4d$, $2s 2p^2 5d$, and $2s 2p^2 3s$, from which single and double excitations can be made, are listed in the column ``extras.'' QED corrections are calculated separately and listed in column ``QED''. 
In addition, the contributions to the transition energies for $^3\!\text{P}_2 -\, ^3\!\text{P}_1$ and
$^3\!\text{P}_1 -\, ^3\!\text{P}_0$ are given in the last two rows. 

Note that the energies of the first three levels seem to have converged fully in terms of the basis set expansion contributions, but not for the last two levels, where higher orbitals give non-negligible contributions. A full convergence for the $i$ and $k$ partial waves would require computational resources beyond current availability. We estimated the contributions of higher-order partial waves by studying the convergence patterns. 
In column ``$>\!k$'', we list contributions to the energies from extrapolations beyond the $k$ partial wave.
Additionally, we include contributions from triple excitations from the $1s^2 2s^2 2p^2$ and $1s^2 2p^4$ configurations to a smaller $12spdf\!g$ basis set under the column ``triples''. 
The final values are given in cm$^{-1}$ in column ``Final.'' The differences between the final values and the experimental values from NIST and Ref.~\cite{PRA} are given in cm$^{-1}$  in the last two columns. 

Finally, we add the frequency-dependent Breit interaction corrections. These corrections were calculated as the expectation value of the symmetrized one-photon exchange operator~\cite{Mann_Johnson1971,Mittleman1972,Grant1976,Tupitsyn2020} using the CI wave functions in a basis set that includes only the single-electron $1s$, $2s$, and $2p$ Hartree-Fock-Dirac orbitals. We note that this contribution is usually considered negligible and was omitted in all of our previous computations. However, there was a clear indication of the missing contributions at the level of 10 cm$^{-1}$ for the $^3\!\text{P}_1$ level. The accuracy of the QED contribution was recently evaluated in Ref.~\cite{Fe2024} and the size of the QED contribution for the transition energies listed in the last two rows in Table~\ref{energies} did not match the size of the discrepancy. The CI contribution clearly converges for these states, indicating that there was another source of the difference. 

We isolated the contribution of the Breit interaction in Table~\ref{energies-breit_comp}, where we list contributions to the energies from the inclusion of the Breit interaction and the frequency-dependent correction. We find that the Breit corrections are very large, a few percent of the total excitation energies. This is due to large cancellations in the total energies of the ground and excited states, leading to 
%low-lying 
optical transitions between them. We note that such a large Breit interaction is common for the highly charged ions of interest in the development of high-precision atomic clocks \cite{HCI1,HCI2}. Our calculations show that frequency-dependent Breit contributes at the level of 1\% of the Breit interaction, enabling us to estimate in which cases frequency-dependent Breit should be computed to achieve the expected accuracy.

The contributions to the $g$-factors for the $^3\!\text{P}_1$, $^3\!\text{P}_2$, and $^1\!\text{D}_2$ states are listed in Table~\ref{$g$-factors}. Non-relativistic $g$-factors calculated using the well-known Land\'e $g$-factor formula are listed in column ``NR'' and relativistic values for a point-like nucleus in column labeled ``Dirac''. Contributions to the $g$-factors from the inclusion of the $h$ and $i$ partial waves are listed in the columns ``$+24h$'' and ``$+24i$'', respectively. 
%The results obtained using this approach in the rows labeled ``NR'' in Table~\ref{$g$-factors}.

We use the operator of the interaction with the external homogeneous magnetic field $\bm{B}$,
$$V_m = \frac{ec}{2} \sum_i(\bm{r_i} \times \bfalpha_i) \cdot \bm{B},$$ where
% $
% \bfalpha=
% \begin{pmatrix}
%      0       & {\bm \sigma} \\
% {\bm \sigma} &     0 
% \end{pmatrix},
% $
%$\bm \sigma$ are the Pauli matrices, 
$e$ is the elementary charge, $c$ is the speed of light, $\bfalpha$ is the vector of the Dirac matrices, and $i$ enumerates the electrons of the ion.
This operator mixes the large and small components of the
wave functions. In this case, negative-energy states also contribute to the $g$-factor. Their
contribution, labeled ``neg.'' in Table~\ref{$g$-factors}, was taken into account using the following expression:
\begin{equation}
    \frac{ec}{\mu_{\rm B}} \frac{1}{J} \sum_{p,n}\frac{\langle p | [\bm{r} \times \bfalpha ]_z | n \rangle }{\varepsilon_p - \varepsilon_n}\langle \hat{a}_n^\dagger \hat{a}_p \Psi | H | \Psi \rangle,
\end{equation}
where $|p\rangle$ and $|n\rangle$ denote the positive- and negative-energy one-electron states with energies $\varepsilon_p$ and $\varepsilon_n$, respectively. $\hat{a}^\dagger$ and $\hat{a}$ are the corresponding creation and annihilation operators~\cite{Zinenko2023}, and $\Psi$ is the wave function of the state with $m_J = J$. 

We note that including negative-energy states was shown to be critical for the computation of the magnetic dipole ac polarizabilities of the Sr clock states \cite{Sr,Sr1}. This is also important for the high-precision calculation of  magnetic dipole transition amplitudes between levels that do not belong to the same fine-structure multiplet, such as the $6s - 7s$ transition in Cs.

The inclusion of QED effects in the Hamiltonian affects the wave functions and is crucial to achieve good agreement with the experiment for the energies (cf. Table~\ref{energies}). However, this plays a minor role for the $g$-factors (see Table~\ref{$g$-factors}). We list the contribution to $g$-factors associated with the QED correction to the wave functions in the column labeled ``QED'' for consistency with the energy table.  A much larger contribution to the $g$-factor comes from the QED corrections to the magnetic moment and can be approximately estimated as an expectation value of the operator
\begin{equation}
    \frac{g_{\rm free}-2}{2}\mu_{\rm B}\sum_i \beta_i \Sigma_{z,i}, 
\end{equation}
where $\beta$ is the Dirac matrix,
% \begin{equation*}
% \beta = 
% \begin{pmatrix}
%      I &  0 \\
%      0 & -I 
% \end{pmatrix}, \quad  \quad  
$\Sigma_z =
\begin{pmatrix}
     \sigma_z  & 0 \\
        0      & \sigma_z 
\end{pmatrix}
$,
% \end{equation*} 
 $\sigma_z$ is the Pauli matrix
and $g_{\rm free} = 2[1+0.5(\alpha/\pi)-0.328478\ldots\times(\alpha/\pi)^2+\ldots]$ is the free-electron $g$-factor.
We label this QED correction ``QED$_\textrm{m}$'' in Table~\ref{$g$-factors} to distinguish this contribution from the QED effect on the wave functions. 
A good agreement between this estimate and the rigorous QED calculation was recently demonstrated in boron-like Ar~\cite{Maison2019}.

A comparison of our final results with those of Ref.~\cite{Verdebout2014} is presented in Table~\ref{$g$-factors}. The agreement is good, with small differences due to accounting for correlations. In Ref.~\cite{Verdebout2014}, the same QED estimate was used, but the contribution of the negative-energy states was neglected. While the accurate treatment of the correlation correction is critical for achieving high accuracy for the energies, 
the QED and relativistic corrections are more important for the $g$-factors. 
%A small difference in the theoretical result for the $^3\!\text{P}_1$ state $g=1.49902$ compared to the experimental value $g=1.499032(6)$ is due to the approximate treatment of the QED corrections and the negative-energy contribution.
The theoretical result for the $^3\!\text{P}_1$ state $g=1.49902$ is in good agreement with the experimental value $g=1.499032(6)$. Further calculations should include the rigorous QED treatment and nuclear recoil corrections. 
%Negative energy contributions to $g$-factors were evaluated for simpler systems. For example, the $g$-factor of lithium-like silicon $^{28}$Si$^{11+}$ was calculated with high precision in Ref~\cite{Si}, where the negative energy contribution to the $1/Z$ term was about a third of the positive energy correlation contribution, but contributing with an opposite sign. Here, we are interested in the $2p$ state rather than $2s$, but such an estimate would give the right order of magnitude to the difference with experiment, $-0.0002$.
